\documentclass[
reprint,
superscriptaddress,
nofootinbib,
nobibnotes,
amsmath,amssymb,
aps,
pra,
showkeys,
]{revtex4-1}

\usepackage{graphicx}
\usepackage{bm}
\usepackage{amsfonts}
\usepackage[normalem]{ulem}
\usepackage{wasysym}
\usepackage{siunitx}
\usepackage{url}
\usepackage{color}

\newcommand*{\diff}{\,\mathrm{d}}

\usepackage{color}   
\usepackage{hyperref}
\hypersetup{
	colorlinks=true,
	pdfborder = {0 0 0.5 [3 3]},
	anchorcolor=black,
	citecolor=green,
	linktoc=all,
	linktocpage=true,
	linkcolor=red,
	linkbordercolor={1 1 0},
}

\begin{document}

\title{Modeling and searching for a stochastic gravitational-wave background\\ from ultralight vector bosons}

\author{Leo Tsukada}
\email{tsukada@resceu.s.u-tokyo.ac.jp}
\affiliation{
	Research Center for the Early Universe (RESCEU), Graduate School of Science, \\The University of Tokyo, Tokyo 113-0033, Japan}
\affiliation{Department of Physics, Graduate School of Science, The University of Tokyo, Tokyo 113-0033, Japan}

\author{Richard Brito}
\affiliation{Dipartimento di Fisica, ``Sapienza'' Universit\`a di Roma \& Sezione INFN Roma1, Piazzale Aldo Moro 5,
00185, Roma, Italy}

\author{William E. East}
\affiliation{Perimeter Institute for Theoretical Physics, Waterloo, Ontario N2L 2Y5, Canada}
\author{Nils Siemonsen}
\affiliation{Perimeter Institute for Theoretical Physics, Waterloo, Ontario N2L 2Y5, Canada}
\affiliation{Department of Physics \& Astronomy, University of Waterloo, Waterloo, ON N2L 3G1, Canada}

\date{\today}

\begin{abstract}
Ultralight bosons, which are predicted in a variety of beyond-Standard-Model
scenarios as dark-matter candidates, can trigger the superradiant instability
around spinning black holes. This instability gives rise to oscillating boson
condensates which then dissipate through the emission of nearly monochromatic
gravitational waves. Such systems are promising sources for current and
future gravitational-wave detectors. In this work, we consider
minimally-coupled, massive vector bosons, which can produce a significantly
stronger gravitational-wave signal compared to the scalar case. We adopt
recently obtained numerical results for the gravitational-wave flux, and
astrophysical models of black hole populations that include both isolated
black holes and binary merger remnants, to compute and study in detail the
stochastic gravitational-wave background emitted by these sources. Using a
Bayesian framework, we search for such a background signal
emitted using data from the first and second observing runs
of Advanced LIGO. We find no evidence for such a signal. Therefore, the
results allow us to constrain minimally coupled vector fields with masses in
the range $\SI{0.8e-13}{\electronvolt}\leq m_b\leq
\SI{6.0e-13}{\electronvolt}$ at 95\% credibility, assuming optimistically
that the dimensionless spin distribution for the isolated black hole
population is uniform in the range $[0,1]$. With more pessimistic
assumptions, a narrower range around $m_b\approx\SI{e-13}{\electronvolt}$ can
still be excluded as long as the upper end of the uniform distribution for
dimensionless black hole spin is $\gtrsim 0.2$.
\end{abstract}

\maketitle

\section{\label{sec:1}Introduction}
	The detection of gravitational waves (GWs) emitted by binary black-hole
	(BBH) and binary neutron stars coalescence
	events~\cite{GW150914,GW151226,O1_catalog,GW170104,GW170608,GW170814,GW170817,O2_catalog,GW190412,Abbott:2020uma,Abbott:2020khf,Abbott:2020tfl,Abbott:2020niy}
	has opened a new era of discoveries with far-reaching implications for
	astrophysics \cite{TheLIGOScientific:2016htt, O2_pop,Barack:2018yly,Abbott:2020gyp} and
	fundamental
	physics~\cite{TheLIGOScientific:2016src,Abbott:2018lct,LIGOScientific:2019fpa,Barack:2018yly,Bertone:2019irm,Abbott:2020jks,CalderonBustillo:2020srq}.
	In the very near future~\cite{Aasi:2013wya}, Advanced
	LIGO~\cite{TheLIGOScientific:2014jea} and Advanced
	Virgo~\cite{TheVirgo:2014hva} will be joined in observing by additional
	detectors, such as KAGRA~\cite{kagra} and LIGO-India \cite{ligo-india},
	and there are plans for a third generation of ground-based
	detectors~\cite{Maggiore:2019uih,Reitze:2019iox}. Together with the
	planned space-based GW detectors LISA~\cite{Audley:2017drz}, and
	pulsar-timing arrays~\cite{Perera:2019sca}, this will allow us to access
	a large range of the GW frequency spectrum.

	A major target for this network of detectors is the detection of a
	stochastic gravitational-wave background (SGWB) produced by the
	incoherent superposition of many sources too faint to be resolved
	individually (see e.g.~\cite{Christensen:2018iqi} for a recent review).
	In the LIGO/Virgo frequency band, one of the most promising targets is
	the background emitted by compact binary coalescences
	(CBCs)~\cite{GW150914_implication,O1_iso,LVC_O2iso,GW170817_implication}.
	Here we consider another possible source for the SGWB that would be
	present if ultralight bosons in certain mass ranges exist in the
	Universe~\cite{brito_short,Tsukada:2019} (see also Ref.~\cite{xilong}).

	The main mechanism responsible for this SGWB is the superradiant
	instability of spinning black holes (BHs) in the presence of massive
	bosons~\cite{Damour:1976kh,Zouros:1979iw,Detweiler:1980uk,Dolan,string_axiverse,Shlapentokh-Rothman:2013ysa,Pani:2012vp,Pani:2012bp,Witek:2012tr,sr_tensor,Endlich:2016jgc,East:2017mrj,East:2017ovw,sr_vector_4,Cardoso:2018tly,East:2018glu,Frolov:2018ezx,Dolan:2018dqv,Baumann:2019eav,Brito:2020lup}.
	The superradiant instability relies on the fact that massive bosons
	with rest mass $m_b$ can form (quasi-)bound states around BHs with
	oscillation frequency $\omega_R\sim m_b c^2/\hbar$, allowing for
	continuous energy extraction whenever $\omega_R$ satisfies the
	\emph{superradiant condition}
	\begin{equation}
		\label{eq:superradiant}
		0<\omega_R<m\Omega_H \ ,
	\end{equation}
	where $m$ is the azimuthal index of the boson field, and $ \Omega_H $
	is the BH's horizon angular velocity (see Ref.~\cite{superradiance} for a
	review). As the system becomes unstable, the boson modes start growing
	exponentially. The
	superradiant instability is most effective when the boson's reduced
	Compton wavelength $\lambdabar\equiv\hbar/(m_b c)$ is comparable to the
	BH's gravitational radius $r_g\equiv 2GM/c^2$, i.e., when
	\begin{align}
		\label{eq:Mmu1}
		m_bc^2&\sim \frac{\hbar c^3}{2GM}\sim\SI{e-12}{\ \electronvolt}\times\left(\frac{M}{\SI{70}{\ M_{\odot}}}\right)^{-1},
	\end{align}
	for a BH with mass $M$. In general, for a given boson rest mass, only for
	a relatively narrow window of BH masses will the superradiant instability
	timescale be sufficiently short in an astrophysical context
	(see e.g. Ref~\cite{superradiance}).

	During the instability phase, the BH spins down, transferring energy and
	angular momentum to the boson field until the point where the
	superradiant condition saturates $\omega_R\sim \Omega_ H$, resulting in
	the formation of an oscillating non-axisymmetric boson ``cloud'' which
	acts as a source of nearly-monochromatic GWs with frequency
\begin{equation}
f_{\rm GW}\sim \omega_R/\pi \sim 484\ \,{\rm Hz}\left(\frac{m_b c^2}{\SI{e-12}{\ \electronvolt}}\right)\,. 							\label{eq:gw_freq}
\end{equation}
	Combining the above equation with Eq.~\eqref{eq:Mmu1}, it follows that
	Advanced LIGO is especially sensitive to the GW emission from bosons with
	$m_b \sim\mathcal{O}(\SI{e-12})$ eV surrounding BHs with masses from
	$\mathcal{O}(10) M_{\odot}$ up to $\mathcal{O}(100) M_{\odot}$. These GWs
	have been shown to be observable with current and future ground-based GW
	detectors in two regimes---a ``resolvable'' regime, in which nearby
	sources can be directly detected~
	\cite{string_axiverse,Arvanitaki_precision,Arvanitaki1,Arvanitaki2,brito,sr_vector_4,
	allsky_sr_method,allsky_sr_search,
	directional_sr_method,Ghosh:2018gaw,directional_sr_CygX1,CW_galactic,Brito:2020lup},
	and an ``unresolvable'' regime, where the incoherent superposition of all
	other sources in the Universe contribute to a
	SGWB~\cite{brito,brito_short,Tsukada:2019,Brito:2020lup}.

	These considerations are especially important given that light boson
	fields in a wide range of masses have been proposed as potential dark
	matter
	candidates~\cite{string_axiverse,dark,Marsh:2015xka,Hui:2016ltb,Irastorza:2018dyq,Baldeschi:1983mq},
	and are predicted in many extensions to the Standard Model of particle
	physics~\cite{Goodsell:2009xc,string_axiverse,Jaeckel:2010ni,dark,Graham:2015rva,Agrawal:2018vin,Irastorza:2018dyq}.
	Prototypical examples include not only the hypothetical QCD
	axion~\cite{Peccei:1977hh,Weinberg:1977ma}, axion-like particles arising
	in string theory scenarios~\cite{string_axiverse}, but also models
	involving ultralight vector fields, such as dark photons as dark matter
	candidates~\cite{dark,Graham:2015rva,Co:2018lka,Agrawal:2018vin,Nakai:2020cfw,Namba:2020kij}, and more generic
	hidden vector fields which also arise as a generic prediction of string
	theory~\cite{Goodsell:2009xc}. Being a purely gravitational effect, the
	superradiant instability and subsequent GW emission from boson clouds
	provide a powerful way to search for such particles and complements more
	conventional searches which normally rely on (non-gravitational)
	couplings of these fields with Standard Model particles.

	Most studies and searches for GW signals from boson clouds in LIGO data
	have so far focused on massive scalar fields. In particular,
	Ref.~\cite{Tsukada:2019} conducted a search for the SGWB model in the
	data from LIGO's first observation run. No signal was found, which
	allowed them to constrain scalar fields with masses in the range
	\SIrange[]{2.0e-13}{3.8e-13}{\electronvolt} in an optimistic scenario.
	Excluding some range of ultralight boson masses in the absence of the
	detection of GW signal requires one to make assumptions about the BH
	population and, in particular, the BH spin distribution. Similarly in this work, we
	will consider several different ways of parameterizing the unknown BH
	population statistics. Searches for (resolvable) continuous GWs emitted
	by individual BH-scalar cloud systems have also been
	conducted~\cite{directional_sr_CygX1, allsky_sr_search, CW_galactic} but
	no signal has been found so far either, suggesting constraints on scalar
	bosons in a similar mass range ($\sim [10^{-13},10^{-12}]$ eV).

	Making use of recent theoretical developments in the understanding of
	superradiant instabilities from vector
	fields~\cite{Frolov:2018ezx,Dolan:2018dqv,Siemonsen:2019ebd}, in this
	paper we extend those results by modeling in detail the SGWB emitted by
	vector fields and searching for this signal in Advanced LIGO's data.
	Compared to the scalar field case, the superradiant instability and GW
	emission timescales for vector fields can be significantly shorter.
	Intuitively, this is because ultralight vector clouds can carry spin
	angular momentum (while scalar clouds can only carry orbital angular
	momentum), and thus vectors form more compact clouds with greater fluxes
	across the BH horizons. As we will show, these faster timescales allow us
	to constrain a wider range of boson masses.

	This paper is structured as follows. In Sec.~\ref{sec:2}, we provide a
	brief overview of the superradiant instability and subsequent GW emission
	by massive vector bosons presented in Ref.~\cite{Siemonsen:2019ebd}. In
	Sec.~\ref{sec:3}, we discuss in detail the predicted SGWB signal from
	vector clouds and compare it to the scalar field case. In
	Sec.~\ref{sec:4}, we present the Bayesian framework that we use to search
	for this background in GW data. Using this framework, in Sec.~\ref{sec:5}
	we study the vector mass range that this search method is sensitive to,
	and explore the capacity of Bayesian model selection to distinguish
	between the background due to the superradiant instability and that due
	to unresolved CBCs. A search using real data is presented in
	Sec.~\ref{sec:O1O2_zerolag}, where we show the range of excluded vector
	masses using data from Advanced LIGO's first and second observing runs.
	Finally, Sec.~\ref{sec:6} summarizes our findings and the implications of
	our results.

	In what follows, we use units $ G=c=1 $ unless otherwise stated.

\section{\label{sec:2}Superradiant instability and GW emission}

	In this section, we briefly review how a vector cloud would
	spontaneously grow around a spinning BH through the superradiant
	instability, eventually saturate, and then dissipate through the emission
	of GWs. The superradiant instability can occur for bosons with spin-0
	(scalar)
	\cite{Damour:1976kh,Zouros:1979iw,Detweiler:1980uk,Dolan,Arvanitaki1,
	Arvanitaki2,kodama_yoshino_2014,brito,brito_short} or spin-1 (vector)
	\cite{ Pani:2012bp, sr_vector_4,East:2017ovw,
	East:2017mrj,East:2018glu,Cardoso:2018tly, Baumann:2019eav} (see also
	Refs.~\cite{sr_tensor,Brito:2020lup} for massive spin-2 fields). The
	qualitative picture is the same in either case, the main difference being
	the generically shorter timescales for the vector field case. We then
	describe our specific model for the GW signal from a vector cloud,
	which is based on Ref.~\cite{Siemonsen:2019ebd}.

	\subsection{Evolution of the Proca cloud}

\begin{figure*}[t]
	\centering
	\includegraphics[width=0.95\textwidth]{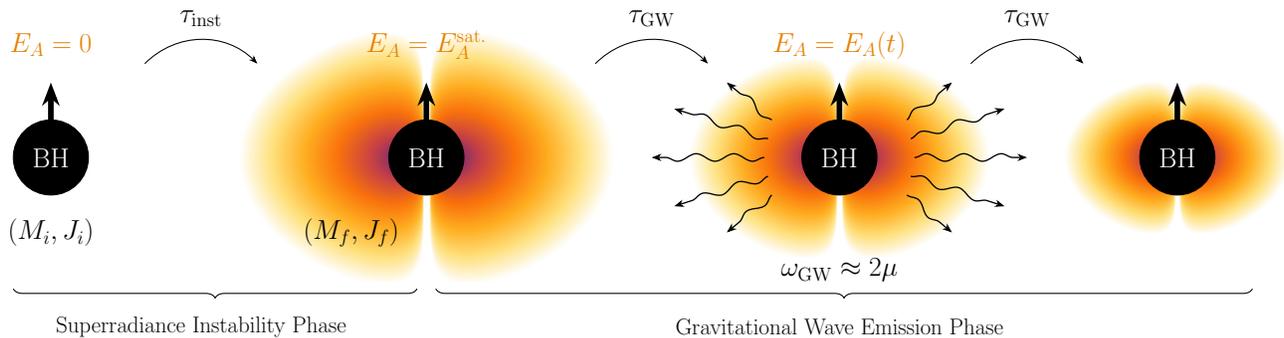}
	\caption{\label{fig:sr_inst_cartoon} A schematic representation of the
	evolution of the superradiance instability and subsequent GW emission.
	Initial (e.g. quantum) fluctuations in the Proca field seed the
	instability, leading to an exponentially growing boson cloud around the
	spinning BH (with growth timescale $\tau_\mathrm{inst}$). The Proca cloud
	grows at the expense of BH angular momentum and mass:
	$M_i-M_f=E_A^\text{sat.}>0$. After saturation, in the GW emission phase
	the cloud slowly decays with timescale $\tau_\mathrm{GW}$ by emitting
	monochromatic gravitational radiation at frequency
	$\omega_\mathrm{GW}\approx 2\mu$ (see also Eq.~\eqref{eq:omega_R_nonR}),
	until the cloud's mass is too small to emit detectable GWs, or an
	unstable higher azimuthal mode begins dominating the dynamics.}
\end{figure*}	
	
	We consider a single (\textit{real}) massive vector $A_\mu$, or Proca
	field, which is minimally coupled, and ignore any coupling with Standard
	Model particles, as well as any non-trivial self-interactions beyond the
	mass term. Around a spinning BH, bound Proca states with oscillation
	frequency $\omega_R$ satisfying Eq.~(\ref{eq:superradiant}) can
	spontaneously grow, exhibiting exponential growth with imaginary
	frequency $\omega_I$. The BH-cloud system is then characterized by three
	distinct timescales: The oscillation timescale
	$\tau_\text{osc.}=\omega_R^{-1}$, the superradiant instability growth
	timescale $\tau_\text{inst}=\omega_I^{-1}$, and the GW emission timescale
	$\tau_\text{GW}$ (defined below). The hierarchy of these timescales,
	$\tau_\text{osc.}\ll\tau_\text{inst}\ll\tau_\text{GW}$, enables us to
	treat the extraction of angular momentum from the spinning BH in a
	quasi-adiabatic form and ignore gravitational radiation during the
	evolution of the superradiant instability~\cite{brito_review}. See
	Fig.~\ref{fig:sr_inst_cartoon} for a cartoon representation of the
	dynamics of the BH-cloud system from the onset of the superradiant
	instability, through saturation and to the GW emission phase. Assuming
	that the instability is triggered by some small initial Proca field
	configuration [e.g. a quantum fluctuation with $\mathcal{O}(1)$ massive
	vector bosons], it grows exponentially with timescale $\tau_\text{inst}$.
	In the limit $M\mu\ll 1$ the typical instability timescales are roughly
	given by (see e.g.~\cite{Baumann:2019eav})
	\begin{eqnarray}
	&&\tau_{\rm inst}^{S} \approx 30\, {\rm days} \left(\frac{M}{10\, M_{\odot}}\right) \left(\frac{0.1}{M\mu}\right)^{9}\hspace{-2pt} \left(\frac{0.9}{\chi_i}\right)\,, \label{eq:tinst_sca}\\
	&&\tau_{\rm inst}^{V}\approx 280\, {\rm s} \left(\frac{M}{10\, M_{\odot}}\right) \left(\frac{0.1}{M\mu}\right)^{7}\hspace{-2pt} \left(\frac{0.9}{\chi_i}\right)\,,\label{eq:tinst_vec}
	\end{eqnarray}
	where the superscripts $S$ and $V$ each stands for the scalar and vector
	field case, respectively.
	While doing so, it extracts angular momentum $\delta J$ from the Kerr BH.
	For linear fluxes across the BH horizon, it follows that $\delta J=m
	\tau_\text{osc.} \delta M$. Hence, we can assume that the BH-cloud system
	moves through a sequence of Kerr spacetimes with decreasing angular
	momentum $J$ and mass $M$ (but \textit{increasing} irreducible mass).
		
	The amplitude of the Proca field $A_\mu$ increases by roughly $180$
	e-folds from the onset of the instability to saturation. As was noted in
	Refs. \cite{East:2017ovw,Herdeiro:2017phl,East:2018glu}, the BH-cloud
	system is well-modeled by the \textit{linear} Proca solution on the
	background of a Kerr BH with mass and angular momentum that slowly
	decreases until the synchronization criterion
	\begin{align}
	\Omega_H(M_f,J_f)=\Omega_H(M_i + \delta M,J_i+\delta J)=\omega_R/m,
	\label{eq:synchcondition}
	\end{align}
    is satisfied and the instability shuts off. Here $M_i$ and $J_i$ are the
    initial BH mass and angular momentum, and $M_f$ and $J_f$ are the final
    BH parameters, and $\Omega_H(M,J)=J/[2M(M^2+(M^4-J^2)^{1/2})]$ is the BH
    horizon frequency. This expression, combined with knowledge of $\omega_R$
    (which has implicit dependence on $M$ and $J$) determines the energy
    $E_A^\text{sat.}=-\delta M$ that is extracted from the BH. Depending on
    the parameters, the cloud can contain up to $\sim 10 \%$ of the original
    BH's mass, while oscillating with frequency $\omega_R$ around the BH.
    This induces strong gravitational radiation that can potentially be
    observed.\footnote{Note, a \textit{complex} massive vector field
    undergoes the same exponential growth as its real counterpart. However,
    if the real and imaginary components of the field are arranged such that
    the resulting stress-energy distribution is axisymmetric (assuming
    similar initial conditions as above), the GW emission is highly
    suppressed compared to the case considered here~\cite{East:2017ovw}.}.
    After saturation, the only dynamical timescales are $\tau_\text{osc.}$
    and $\tau_\text{GW}$. Since the GW power is proportional to the square of
    the cloud energy, $\dot{E}_\text{GW}\propto E_A^2$, the cloud energy
    reduces as
	\begin{align}
	E_A(t)=\frac{E^\text{sat.}_A}{1+t/\tau_\text{GW}}, & & \tau_\text{GW}=\frac{E^\text{sat.}_A}{\dot{E}_\text{GW}(t=t_\text{sat.})}.
	\label{eq:t_gw}
	\end{align}

	The GW power for vector clouds is to be contrasted with the scalar
	field case, for which the GW power is much smaller~\cite{Siemonsen:2019ebd}. This
	difference in the GW power translates in a large difference in the
	typical GW emission timescale, of which the non-relativistic estimates
	(i.e. $M\mu\ll 1$) roughly read:
	\begin{eqnarray}
	&&\tau_{\rm GW}^{S} \approx 10^{5}\, {\rm yr} \left(\frac{M}{10\, M_{\odot}}\right) \left(\frac{0.1}{M\mu}\right)^{15}\hspace{-2pt} \left(\frac{0.5}{\chi_i-\chi_f}\right)\,, \label{eq:tgw_sca}\\
	&&\tau_{\rm GW}^{V}\approx 8\, {\rm days} \left(\frac{M}{10\, M_{\odot}}\right) \left(\frac{0.1}{M\mu}\right)^{11}\hspace{-2pt} \left(\frac{0.5}{\chi_i-\chi_f}\right)\,,\label{eq:tgw_vec}
	\end{eqnarray}
	where again the superscripts $S$ and $V$ each stands for the scalar and vector
	field case, and $\chi_i$ and $\chi_f$ stand for the BH spin at birth and
	the end of the instability phase, respectively. Comparing these to the
	instability timescales, Eqs.~\eqref{eq:tinst_sca} and
	\eqref{eq:tinst_vec}, it clearly follows that $\tau_{\rm GW} \gg
	\tau_{\rm inst}$, and hence during the exponential evolution of the Proca
	cloud, the GW emission can be neglected. This is true even
	beyond the non-relativistic regime~\cite{Siemonsen:2019ebd}.

	\subsection{Proca-BH bound states}
	In our approach, which follows closely the Proca mode analysis in
	Refs.~\cite{Siemonsen:2019ebd,Dolan:2018dqv}, we use BH perturbation
	theory to compute the bound Proca states.  The underlying field
	equations are
	\begin{align}
	\nabla_\alpha F^{\alpha\beta} = \mu^2 A^\beta,
	\label{eq:ProcaFEs}
	\end{align}
	where $F_{\alpha\beta} = 2\nabla_{[\alpha}A_{\beta]}$ is the Proca field
	strength tensor, $\mu=m_b/\hbar$, and geometric quantities are computed
	using the Kerr metric of a spinning BH. Lunin discovered a separation
	ansatz in Ref.~\cite{Lunin:2017drx} for the Maxwell equations in Kerr
	spacetimes (and generalizations thereof) that makes a direct
	reconstruction of the 4-potential trivial after solving the respective
	second order ordinary differential equations (ODEs).\footnote{This can be
	contrasted with the well-established Teukolsky formalism
	\cite{Teukolsky:1973ha} that provides only a single polarization and
	requires an elaborate reconstruction mechanism to construct the
	4-potential.} In Refs.~\cite{Frolov:2018ezx, Krtous:2018bvk}, this ansatz
	was shown to separate the massive vector field equations in Kerr
	spacetimes (which are not separable in the Teukolsky formalism). The
	vector potential ansatz takes the form
	\begin{align}
		A^\mu = B^{\mu\nu}\nabla_\nu Z, & & Z=e^{-i\omega t + im \varphi} R(r)S(\theta),
	\end{align}
	where $B^{\mu\nu}$ is a polarization tensor constructed from the hidden
	symmetries of the Kerr-NUT-(A)dS family of spacetimes (see
	Ref.~\cite{Dolan:2018dqv} for the explicit form used here). With this
	ansatz, Eq.~\eqref{eq:ProcaFEs} reduces to two separated ODEs
	parameterized by the vector boson mass $M\mu$ and dimensionless BH spin
	$\chi=J/M^2$, and coupled only by the respective separation constants:
	the azimuthal mode number $m\geq 1$, the overtone number $\hat{n}\geq 0$,
	the polarization state $S\in\{-1,0,1\}$, and the real and imaginary parts
	of the frequency $\omega=\omega_R+i\omega_I$.

	In the non-relativistic limit, i.e., if $M\mu \ll 1$, \cite{Pani:2012vp, Pani:2012bp, Cardoso:2018tly, sr_vector_4, Baumann:2019eav}
	\begin{align}
			\omega_R = & \ \mu \left(1-\frac{\mu^2 M^2}{2(|m|+\hat{n}+S +1)^2} \right) + \mathcal{O}[(M\mu)^4]\label{eq:omega_R_nonR}\\
			\omega_I = & \ C_{m,\hat{n},S}(J,M,\omega_R) (M\mu)^{4m+2S+5}(\omega_R-m\Omega_H).
			\label{eq:t_inst}
		\end{align}
	where $C_{m,\hat{n},S}(J,M,\omega_R)$ is a set of coefficients
	\cite{Cardoso:2018tly, sr_vector_4, Baumann:2019eav}, and the modes with
	$S=-1$ polarization, $\hat{n}=0$, and lowest value of $m$ that satisfies
	$\omega_R<m\Omega_H$ are the most unstable (fastest growing). Here we
	focus on such modes and restrict to $m=1$ and 2.\footnote{ We note that,
	as shown in Ref.~\cite{Siemonsen:2019ebd}, there are regions of the
	parameter space where the fundamental mode ($\hat{n}=0$) grows slower
	than one or more of the overtones for $m\geq 2$ in the relativistic
	regime. This raises the possibility of several modes being populated
	simultaneously, generating a unique beating GW signal. We do not consider
	such signals here since they will have a small contribution to the
	stochastic GW background.
	} 
Note that in the non-relativistic limit, the fastest growing vector mode
corresponds to a hydrogen-like cloud with orbital number $\ell=0$ spatial
dependence, but with $j=m=1$, due to the spin contribution, where $j$ is the total angular momentum number~\cite{sr_vector_4}. In contrast, the
fastest growing scalar mode has $\ell=1$, and therefore the cloud sits farther
from the BH due to the centrifugal barrier. Hence, in the vector case the
relative flux across the BH horizon is greater, leading to faster superradiant
growth, and the cloud is more compact, leading to greater gravitational
radiation.

	We additionally restrict ourselves to GWs emitted by BH-cloud systems in
	the saturated state (again due to the hierarchy of timescales mentioned
	above). This implies that the angular momentum dependency of $\omega$ can
	be removed by solving for $J_\text{sat.}$ using
	Eq.~\eqref{eq:synchcondition}. We thus need only the real frequency
	$\omega_R$, and can ignore the instability timescale. (For a detailed
	analysis of the growth rates as a function of Proca mass and BH spin in
	the relativistic regime, see
	Refs.~\cite{East:2017mrj,Cardoso:2018tly,Frolov:2018ezx,Dolan:2018dqv,
	Siemonsen:2019ebd}.) We use the numerical data in
	Ref.~\cite{Siemonsen:2019ebd}, obtained from solving the Proca field
	equations as described above for the marginally unstable modes, to fit
	the following functional form, which already includes the leading-order,
	non-relativistic behavior:
	\begin{align}
		\label{eq:omega_R}
		\frac{\omega_R^\text{sat.}}{\mu} = 1 - \frac{(M\mu )^2}{2m^2} + \sum_{\alpha=4}^{10} c^{m}_{\alpha}(M\mu)^\alpha
	\end{align}
	where the fitted coefficients are given in Table~\ref{tab:omega_R_coef} (see
	also Appendix A in Ref.~\cite{Siemonsen:2019ebd} for fits covering the whole
	parameter space).
	\begin{table}[b]
	\centering
	\begin{tabular}{c|ccccccc}
	\hline
	$ m\backslash\alpha $ & 4 & 5 & 6 & 7 & 8 & 9 & 10\\\hline\hline
	1 & $-2.56$ & $13.85$ & $-97.65$ & $349.53$ & $-615.29$ & $532.55$ & $-183.08 $\\\hline
	2 & $-0.076$ & $0.0071$ & $0.029$ & $-0.051$ & $0.14$ & $-0.12$ & $0.034$\\\hline
	\end{tabular}
	\caption{The coefficients, $c_\alpha^m$, defined in Eq.~\eqref{eq:omega_R}, of the higher order terms in $\omega_R$ for the $m=1$ and $m=2$ mode respectively.}\label{tab:omega_R_coef}
	\end{table}
		This value is used to compute the saturation energy and angular momentum of the cloud,
	as well as the GW frequency, as described below.

	\subsection{Gravitational Radiation}
	In order to compute the GW power from the boson cloud, we use the
	stress-energy tensor calculated from the Proca field solutions described
	above, and numerically solve the Teukolsky equation for the GW
	perturbations with this as a source. See Ref.~\cite{Siemonsen:2019ebd}
	for details. Since the stress-energy is quadratic in the field, the
	(angular) frequency of the GW radiation is $\omega_\text{GW}=2\omega_R$.
	The angular dependence of the GWs has spheroidal harmonic components with
	azimuthal number $\pm 2m$, and is dominated by the $\ell=2m$
	contribution, though higher $\ell$ components can be significant in the
	relativistic and high-spin regime (and are included in our calculation of the power).
	Because the GW energy flux scales as $\dot{E}_\text{GW}\propto E_A(t)^2$
	in this treatment, we can phrase these results in terms of the
	mass-rescaled (dimensionless) GW power
	$\dot{\tilde{E}}_\text{GW}=\dot{E}_\text{GW}\times(M/E_A)^2$. In the
	relativistic regime---in particular for $M\mu>0.05$ if $m=1$, and $\mu
	M>0.67$ if $m=2$---we use the following polynomial fitting function for
	convenience
	\begin{align}
	\dot{\tilde{E}}_\text{GW}^{m}=\sum_{\alpha=0}^{N_m} d^m_\alpha\left(M\mu\right)^\alpha,
	\label{eq:GWpower}
	\end{align}
	where the respective coefficients,
	determined from the numerical data of Ref.~\cite{Siemonsen:2019ebd},
	are given in Table~\ref{tab:dE_dt_coef}.
	\begin{table}[b]
	\footnotesize
	\begin{tabular}{c||cccccccc}
	\hline
	$ m\backslash\alpha $ & 0 & 1 & 2 & 3 & 4 & 5 & 6\\\hline
	1 & $-9.6\times 10^{-6}$ & $-0.000064$ & $0.018$ & $-0.27$ & $2.36$ & $-12.8$ & $41.5$ \\\hline
	2 & $-0.00014$ & $-0.019$ & $0.080$ & $0.00011$ & $1.00$ & $-1.95$ & $-9.31$\\\hline\hline
	\end{tabular}
	\begin{tabular}{c||ccccccccc}
	$ m\backslash\alpha $ & 7 & 8 & 9 & 10 & 11 & 12 & 13 & 14&\\\hline
	1 & $-76.9$ & $70.0$ & $-15.9$ & $-10.2$ & - & - & - & - &\\\hline
	2 & $60.5$ & $-165.1$ &  $275.5$ & $-304.2$ & $-224.6$ & $-107.1$ & $29.9$ & $-3.72$ & \\\hline
	\end{tabular}
	\caption{The coefficients, $d^m_\alpha$, for the GW power ansatz in Eq.~\eqref{eq:GWpower}, fitted against the numerical data of Ref.~\cite{Siemonsen:2019ebd}. Here, $N_{m=1}=10$ and $N_{m=2}=14$.}\label{tab:dE_dt_coef}
	\end{table}
	In order to extrapolate our results to the non-relativistic limit, we
	use the following expressions
	\begin{align}
	\begin{aligned}
	\dot{\tilde{E}}_\text{GW}^{m=1}= & \ 1.3\times10^{-12}\left(\frac{M\mu}{0.05}\right)^{10}, & & M\mu\leq 0.05, \\
	\dot{\tilde{E}}_\text{GW}^{m=2}= & \ \frac{ (M\mu)^{14}}{1.0\times10^{5}} + 6.4\times10^{-4} (M\mu)^{16}, & & M\mu\leq 0.67.
	\label{eq:dE_dt}
	\end{aligned}
	\end{align}
	The exponent of the lowest order term in the respective expressions was
	chosen to match that of the analytic calculation in
	Ref.~\cite{sr_vector_4}, but the
	coefficients were determined by fitting against the numerical data of
	Ref.~\cite{Siemonsen:2019ebd} up to $M\mu\approx 0.1 (0.7)$ for $m=1 (2)
	$ respectively (since even in non-relativistic limit one expects leading
	order corrections to the flatspace results in Ref.~\cite{sr_vector_4}).
	Because the GW power is heavily suppressed in this regime, our results
	are not overly sensitive to how this extrapolation is done.

	Finally, the total GW energy emitted over a time $\Delta t$ is given by
	\begin{align}
	\label{eq:E_gw}
	E_\mathrm{GW}=\int_{t=0}^{\Delta t}\diff  t\frac{\diff  E_\mathrm{GW}}{\diff t}=\frac{E_A^\text{sat.}\Delta t}{\Delta t+\tau_\mathrm{GW}}.
	\end{align}
	For our purposes, we will define the signal duration to be the lifetime
	of each BH, namely $\Delta t = t_0 - t(z_f)$, where $t_0$ is the age of
	the Universe, $t_0\approx 13.8$ Gyr, and $t(z_f)$ is the cosmic time at
	the redshift of the BH formation. For all cases of interest, the
	instability timescale is much smaller than the BH lifetime, so we neglect
	the small delay between BH formation and saturation of the superradiant
	instability. For BHs whose age is comparable to the instability
	timescale, the overestimated GW radiation is negligibly small in any
	case, and therefore this approximation does not affect our estimate of
	the overall energy density $\Omega_\mathrm{GW}$. The SGWB is then
	determined from summing over the energy emitted by each BH-cloud system
	over the population of spinning BHs, as we describe in the next section.

\section{\label{sec:3}Modeling the stochastic background}
	In this section, we describe our SGWB model from the whole population of
	BH-cloud systems. The model follows the construction in
	Refs.~\cite{brito_short,Tsukada:2019}, where the superradiant instability
	of ultralight scalar bosons was considered. In particular, we will see
	that some differences between the background emitted by scalar and vector
	clouds arise, mainly due to the larger GW power emitted by vector bosons
	when compared to the scalar case.

	\subsection{General formulation}
	Under the assumptions that the SGWB is (a) isotropic, (b) unpolarized,
	(c) stationary, and (d) Gaussian, the background spectrum can be described in terms of the GW
	energy density per logarithmic frequency interval. This can be computed
	by integrating the SGWB spectrum from individual sources over the
	entire population~\cite{Phinney:2001di},
	\begin{align}
		\begin{aligned}
			\Omega_\mathrm{GW}(f)&\equiv\frac{1}{\rho_c}\frac{\mathrm{d}\rho_{\mathrm{GW}}}{\mathrm{d}\ln(f)}\\
			&=\frac{f}{\rho_c}\int\diff z\frac{\diff t}{\diff z}\int\diff \boldsymbol{\theta}p(\boldsymbol{\theta})R(z; \boldsymbol{\theta})\frac{\diff E_s}{\diff f_s}(\boldsymbol{\theta}).
			\label{eq:omega_gw}
		\end{aligned}
	\end{align}
	Here $ \rho_c $ is the critical energy density required to have a
	spatially flat Universe, $ R(z; \boldsymbol{\theta}) $ is the event rate
	of GW emission per unit comoving volume per unit \textit{source frame}
	time, and $p(\boldsymbol{\theta}) $ is the multivariate probability
	distribution of the source parameters $ \boldsymbol{\theta} $. Since the
	individual sources emit GWs with nearly constant frequencies, the energy
	spectrum from individual signals can be approximated by
	\begin{align}\label{eq:spectrum}
		\frac{\diff E_\mathrm{s}}{\diff f_s}\approx E_\mathrm{GW}\delta(f(1+z)-f_0)
	\end{align}
	where $ f_0=\omega_R/\pi $ [computed using Eq.~\eqref{eq:omega_R}] and $
	E_\mathrm{GW} $ is given by Eq.~\eqref{eq:E_gw}. Note that, due to the
	cosmological redshift $z$, the observed frequency $f$ is related to the
	source frame frequency $f_s$ such that $f=f_s/(1+z)$.

	\subsection{BH population models}
		To compute the stochastic background, we consider two possible BH
		formation channels: isolated extra-galactic BHs, and BBH merger
		remnants. We treat their contribution to the total GW energy density
		independently. Importantly, we do not consider the galactic BH
		population (e.g.\ as recently described in Ref.~\cite{CW_galactic}),
		as our method to search for a SGWB (described in Sec.~\ref{sec:4}) is
		optimized for a Gaussian distributed signal. The signal emitted from
		galactic BHs is expected to add a mostly non-Gaussian and
		non-isotropically distributed component to the stochastic
		background.\footnote{We should also note that, for a given boson
		mass, the GW signals emitted by the galactic population would tend to
		accumulate in a very narrow frequency window around $\omega_R$
		[see Eq.~\eqref{eq:gw_freq}]~\cite{CW_galactic}, unlike the extra-galactic
		component which should be spread over a broader range of frequencies
		due to the cosmological redshift. Our search method is better suited
		for signals that emit in a broad range of frequencies.} This
		non-Gaussian component would typically be vetoed in the process of
		data conditioning in the SGWB search pipeline we use \cite{veto}. Studying the
		specific features of this component goes beyond the scope of this
		paper, but adding it to the search pipeline is certainly an important
		addition for future work and could make the constraints we here
		present even stronger.

		Let us then briefly explain the prescription for each channel (see Ref.~\cite{Tsukada:2019} for more details). In the isolated BH channel, Eq.~\eqref{eq:omega_gw} can be written as
		\begin{align}
			\label{eq:iso_omega_gw}
			\Omega_\mathrm{GW}^\mathrm{iso}(f)=\frac{f}{\rho_c}\int\diff z\frac{\diff t}{\diff z}\int\diff M\diff\chi p(\chi)\frac{\diff \dot{n}}{\diff M}\frac{\diff E_s}{\diff f_s}\,,
		\end{align}
		where $\frac{\diff \dot{n}}{\diff M}$ is the BH formation rate per BH
		mass, which we construct following Ref.~\cite{Tsukada:2019} using a BH
		mass function that spans masses in the range $[3-60]M_\odot$. Since not
		much is known about the spin distribution at birth of isolated
		BHs\footnote{Some predictions for the natal BH spin distribution can be
		found in Ref.~\cite{Belczynski:2017gds} (see their Figs.~1 and 2), where
		it is shown that the BH spin distribution depends very strongly on the
		assumed model for the angular momentum transport in the progenitor
		stars.}, for the probability density of the natal BH spin $ \chi $ we
		assume a uniform distribution
		\begin{align}\label{eq:spin_dist}
			p(\chi) =
			\begin{cases}
				0&(\chi<\chi_\mathrm{ll}, \chi_\mathrm{ul}<\chi)\\
				\frac{1}{\chi_\mathrm{ul}-\chi_\mathrm{ll}}&(\chi_\mathrm{ll}\leq\chi\leq\chi_\mathrm{ul}),
			\end{cases}
		\end{align}
		where $ \chi_\mathrm{ll}$ and $\chi_\mathrm{ul} $ are the lower and
		upper limit of the distribution, respectively. Given that these
		limits in the natal spin distribution of isolated BHs are extremely
		uncertain, for simplicity when searching for this SGWB, we will
		parametrize the distribution in two different ways: (a) leave the
		lower limit $\chi_\mathrm{ll}$ as a free parameter, but fix
		$\chi_\mathrm{ul}=1$; (b) leave the upper limit $\chi_\mathrm{ul}$ as
		a free parameter, but fix $\chi_\mathrm{ll}=0$. In the remainder of
		the text, we will denote these parametrizations as the
		$\chi_\mathrm{ll}$ and $\chi_\mathrm{ul}$ parametrizations,
		respectively. In general, the first case allows for a larger
		amplitude for the background than the second case, since it ensures a
		population of BHs born with high-spin, from which it is possible to
		extract more energy through the superradiant instability. Different
		choices for $ \chi_{\mathrm{ll,ul}}$ can significantly affect the
		background spectrum. Hence, as we will show in
		Sec.~\ref{sec:O1O2_zerolag}, constraints on the vector boson mass
		obtained when searching for such background in LIGO data crucially
		depend on the parametrization one uses.

	For the BBH merger remnant channel, Eq.~\eqref{eq:omega_gw} reduces to
	\begin{align}
		\begin{split}
			\label{eq:bbh_omega_gw}
			\Omega_\mathrm{GW}^\mathrm{rem}(f)=\frac{f}{\rho_c}&\int\diff z\frac{\diff t}{\diff z}\\
			\times&\int\diff \boldsymbol{m}\diff \chi p\,(\boldsymbol{m})R_m(z;\boldsymbol{m})p(\chi)\frac{\diff E_s}{\diff f_s},
		\end{split}
	\end{align}
		where $\boldsymbol{m}$ denotes the component masses of the BBH
		system, $p(\boldsymbol{m})$ is the component mass distribution, and $
		R_m(z; \boldsymbol{m}) $ is the BBH merger rate density for a given $
		\boldsymbol{m}$ and cosmological redshift $ z $. Compared to the spin
		distribution of isolated BHs, the spin distribution $ p(\chi) $ for
		this channel can be more easily constrained, using measurements of
		the spin of remnant BHs observed by Advanced LIGO and Virgo
		\cite{O1_catalog,GW150914,GW151226,GW170104,GW170608,GW170814,O2_catalog,GW190412,Abbott:2020tfl}.
		For a population of merging BHs dominated by near-equal mass BHs that are not rapidly spinning, as
		the majority of the observations made so far suggest, the spin
		magnitude of the final remnant BHs is clustered around $
		0.7$~\cite{Gerosa:2017kvu,Fishbach:2017dwv}. Therefore, for
		simplicity, we assume that all the remnant BHs initially have $
		\chi=0.7$, that is
	\begin{align}
		\label{eq:rem_pchi}
		p(\chi) = \delta(\chi-0.7).
	\end{align}

		To model the BBH merger rate, we follow the prescription described in
		Ref.~\cite{GW150914_implication,GW170817_implication}, calibrating it
		with the local merger rate inferred from the BBHs detected in the
		first two observing runs\footnote{During the writing of this paper,
		the BBH merger rate was updated based on GWTC-2
		\cite{Abbott:2020gyp},
		$\mathcal{R}_\mathrm{BBH}=23.9\mathrm{Gpc}^{-3}\mathrm{yr}^{-1}$.
		Although we do not use this new rate estimate in the search
		presented here, the update wouldn't change the detectability of the
		signal model significantly, as the contribution from the BBH remnant
		population is mostly subdominant.} of Advanced LIGO and Virgo. We
		adopt the rate estimated in Ref.~\cite{O2_catalog}, in particular the
		one derived from the BH mass function with a power law distribution,
		such that
	\begin{align}
		\label{eq:rateNorm}
		\int p(\boldsymbol{m}) R_m(z=0;\boldsymbol{m})\diff\boldsymbol{m}  = 56~\mathrm{Gpc}^{-3}\mathrm{yr}^{-1}.
	\end{align}

		The two assumptions made above contribute to a systematic uncertainty in
		the prediction of the energy density spectrum,
		$\Omega^\mathrm{rem}_\mathrm{gw}(f)$. However, as we will show in the
		next subsection, the contribution from the BBH remnant channel is
		subdominant (compared to the isolated BH channel) for the range of vector
		masses to which current GW detectors are sensitive. Therefore, this
		uncertainty does not affect the results from our search, and hence, the
		constraints on the vector boson mass.

	\subsection{Total background model}
\begin{figure}[h]
		\centering
		\includegraphics[width=\linewidth]{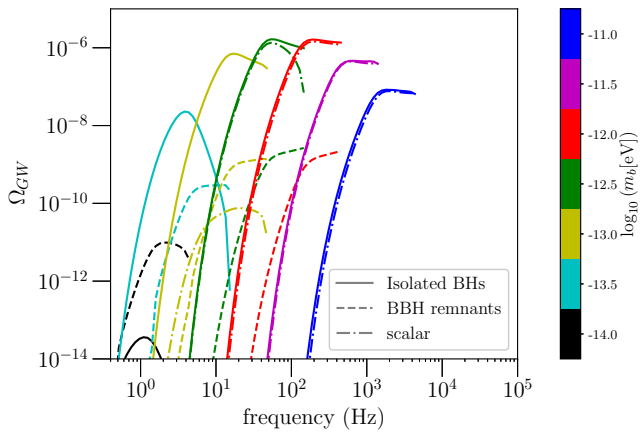}
		\caption{
                Contribution of different BH formation channels for the total
                background spectrum $\Omega_\mathrm{GW}$. Solid curves
                correspond to the spectrum from the isolated BH channel with
                different boson masses represented by the color bar. assuming
                an uniform distribution for the initial BH spin $\chi\in[0,
                1]$, whereas dashed curves show the spectrum due to the BBH
                merger remnant channel. For comparison, we also show the
                total energy spectra that scalar bosons with mass
                $m_b\geq\SI[]{e-13.5}{\ \electronvolt}$ would give rise to
                (dash-dotted lines), including both BH formation channels,
                and assuming the same BH mass and spin distribution as in the
                vector case.
                }
		\label{fig:omega_gw_channels}
	\end{figure}

		We derive the total background by summing over the contributions from the two channels, namely
		\begin{align}
			\label{eq:SI_sgwb}
			\Omega_\mathrm{GW}(f) = \Omega_\mathrm{GW}^\mathrm{iso}(f) + \Omega_\mathrm{GW}^\mathrm{rem}(f),
		\end{align}
		where the superscripts represent each of the isolated BH and BBH
		merger remnant populations defined by Eqs.~\eqref{eq:iso_omega_gw}
		and~\eqref{eq:bbh_omega_gw}, respectively.
		Fig.~\ref{fig:omega_gw_channels} compares the contribution to the
		total energy density spectrum from each of these two channels, where
		we assumed a uniform distribution for the natal BH spin $\chi\in[0,
		1]$ in the isolated BH channel. As one can see, the isolated BH
		channel (solid lines) dominates over the BBH merger remnant channel
		(dashed lines) for $m_b \gtrsim\SI{e-13.5}{\electronvolt}$,
		corresponding to frequencies $\gtrsim \SI{10}{\hertz}$. Since current
		GW detectors are mainly sensitive in this frequency range (see
		Fig.~\ref{fig:omega_gw_PIs}), the detectable SGWB from ultralight
		vector bosons is expected to be dominated by the isolated BH channel.

		For completeness, in Fig.~\ref{fig:omega_gw_channels} we also compare
		the backgrounds emitted by vector bosons against
		those produced by scalar bosons
		(dash-dotted lines) considered in
		Refs.~\cite{brito_short,Tsukada:2019}. For $m_b >
		\SI{e-12.5}{\electronvolt}$, both cases predict almost identical
		spectra. This is because in both cases, the typical instability and
		GW emission timescales for these boson masses are sufficiently short,
		such that for most BHs that become superradiantly unstable, almost
		all the energy in the cloud is dissipated away in GWs within the
		lifetime of the BHs we consider in our population models. Since the
		total amount of energy that can be extracted through the superradiant
		instability is nearly independent of the boson spin, it therefore
		follows that total energy emitted in GWs by the whole BH population
		should be almost independent of the boson spin for boson masses $m_b
		> \SI{e-12.5}{\electronvolt}$.

		On the other hand, for $m_b \lesssim \SI{e-12.5}{\electronvolt}$,
		there can be significant differences between the background predicted
		in the scalar field case, and the one due to a vector field. For $m_b
		= \SI{e-12.5}{\electronvolt}$, this difference is more pronounced at
		larger frequencies, because of the fact that the GWs emitted at
		higher frequencies are typically sourced by BHs at smaller redshifts [see Eq.~\eqref{eq:spectrum}], and hence BHs with smaller lifetimes. Since the
		typical GW emission timescale for scalar bosons is larger than that for
		vector bosons, the total energy emitted in GWs for those BHs tends to
		be smaller for the scalar case than the vector case. Hence, the
		difference at higher frequencies seen for $m_b \lesssim
		\SI{e-12.5}{\electronvolt}$. For lighter boson masses ($m_b <
		\SI{e-12.5}{\electronvolt}$), the amplitude of the background
		predicted in the scalar field case tends to be much smaller than for
		vector fields, because for those boson masses the coupling $M\mu$ is
		very small for most of the BHs in our population [see
		Eq.~\eqref{eq:Mmu1}] and therefore the total energy emitted in GWs
		over the lifetime of those BHs is typically much smaller for the
		scalar field case. In particular, for scalar fields with masses $m_b
		= \SI{e-14}{\electronvolt}$ and $\SI{e-13.5}{\electronvolt}$, the
		typical GW power for the systems in our BH population models is so
		small that the background spectra do not even appear in
		Fig.~\ref{fig:omega_gw_channels}.

		In addition to the comparison between the scalar and vector cloud
		models, Fig.~\ref{fig:omega_gw_channels} also shows a significant
		difference between the two BH formation channels. The spectra from
		the BBH merger remnant channel for vector boson masses $m_b \geq
		\SI{e-11.5}{\ \electronvolt}$ are strongly suppressed because such
		vector fields tend to induce strong superradiant instabilities only
		in lighter BHs ($M\lesssim \SI{10}{\ M_\odot}$), which are less
		likely to be produced by this channel. In addition, one can notice
		from Fig.~\ref{fig:omega_gw_channels} that for $m_b=\SI{e-14}{eV}$,
		the background due to the BBH remnant channel predicts an higher
		amplitude compared to the background induced by the isolated BH
		channel. This is to be contrasted with what happens for heavier
		vector bosons, where the opposite is true. This can be explained from
		the fact that the BHs formed through the isolated BH channel
		(typically $\mathcal{O}(10M_\odot)$) are on average much lighter than
		the ones formed through the BBH merger remnant channel (typically
		$\sim\mathcal{O}(50M_\odot)$ or more), such that for
		$m_b=\SI{e-14}{eV}$ the GW emission timescale is typically much
		larger for the isolated BH channel compared to the BBH merger remnant
		channel (see Eq.~\eqref{eq:tgw_vec}). Therefore, for this boson mass,
		BBH merger remnants tend to radiate more energy within the lifetime
		of the BHs, leading to the strong suppression of the overall
		amplitude for the isolated channel with respect the merger remnant
		channel that we see in Fig.~\ref{fig:omega_gw_channels}. This
		hierarchical flip between the two BH formation channels also occurs
		in the scalar field case, but occurs at $m_b\sim\SI{e-13}{eV}$ due to
		the larger emission timescale for scalar fields (see
		Ref.~\cite{Tsukada:2019}).
		
\begin{figure}
		\centering
		\includegraphics[width=\linewidth]{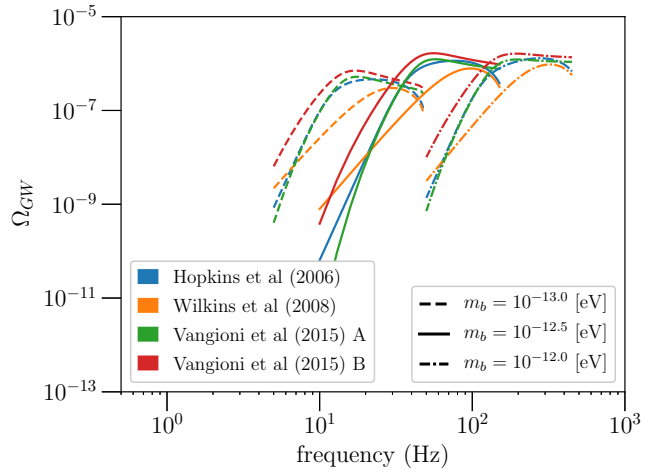}
		\caption{\label{fig:vector_omega_sfrs} Energy density spectra
		assuming different SFR models. Here, we adopt the following four SFR
		models: Hopkins \textit{et al.} 2006 \cite{sfr_hopkins} (blue),
		Wilkins \textit{et al.} 2008 \cite{sfr_wilkins} (yellow), and two
		models from Vangioni \textit{et al.} 2015 \cite{SFR} (green and red).
		Vangioni \textit{et. al.} 2015 A/B represent different ways of
		calibrating the nominal SFR function, i.e. model A calibrates it to
		the observational rate of gamma-ray bursts \cite{sfr_cal_kistler} and
		model B calibrates it to observations of luminous galaxies
		\cite{sfr_cal_Behroozi, sfr_cal_Bouwens}. The different linestyles
		indicate the three vector masses, \SIrange{e-13}{e-12}{eV}. The
		contributions from both BH populations (isolated and merger remnant)
		are included under the assumption that the isolated BHs' spins are
		uniformly distributed over $ \chi \in [0, 1] $.}
	\end{figure}

		Lastly, in Fig.~\ref{fig:vector_omega_sfrs} we study how
		astrophysical uncertainties related to the choice of the BH
		population models impact the the SGWB spectra. More specifically,
		while the BH mass function and local BBH merger rate we adopt are
		motivated by the theoretical and observational constraints as
		described in Ref.~\cite{Tsukada:2019}, there are currently several
		models for the cosmic star formation rate (SFR). Fig.~
		\ref{fig:vector_omega_sfrs} shows how the energy density spectra
		changes assuming four different SFR models: Hopkins \textit{et al.}
		2006 \cite{sfr_hopkins} (blue), Wilkins \textit{et al.} 2008
		\cite{sfr_wilkins} (yellow), and two models from Vangioni \textit{et
		al.} 2015 \cite{SFR} (green and red). Vangioni \textit{et al.} 2015
		A/B represent different ways of calibrating the nominal SFR function.
		Model A calibrates it to the observational rate of gamma-ray bursts
		\cite{sfr_cal_kistler}, whereas model B calibrates it to observations
		of the luminous galaxies \cite{sfr_cal_Behroozi, sfr_cal_Bouwens}. We
		note that our BH population modeling implicitly assumes the SFR model
		of Vangioni \textit{et. al.} 2015 A. The contributions from both the
		isolated BH and BBH remnant channels are included under the
		assumption of isolated BH spin uniformly distributed over $ \chi \in
		[0, 1] $. We find that over the boson mass range of interest,
		\SIrange{e-13}{e-12}{eV}, the uncertainties in the SFR would bring an
		astrophysical uncertainty of approximately a factor of 10 or less.
		This is typically much smaller than the uncertainty related to the
		unknown BH spin distribution. The SGWB spectrum predicted with the
		SFR model of Vangioni \textit{et al.} 2015 A lies between the other
		SFR models, and thus the model we use in our analysis can be
		considered an intermediate scenario given the astrophysical
		uncertainty.

	\subsection{\label{sec:higher_order_mode} Impact of modes with $m>1$}

		The results shown above only take into account the mode with the
		smallest instability timescale, i.e. $m=1$. However, the superradiant
		instability occurs for any azimuthal number $m$, as long as the
		superradiance condition Eq.~\eqref{eq:superradiant} is satisfied. For
		some values of the boson mass and BH parameters, this condition will
		only be satisfied for $m>1$ (either due to the BH's properties at
		birth, or because the BH has been spun down by the $m=1$ mode growing
		to saturation), making these modes relevant. As can be seen from
		Eqs.~\eqref{eq:t_inst} and~\eqref{eq:dE_dt}, the instability and GW
		emission timescales increase with $m$ and therefore, in general, we
		expect the dominant contribution to the background to come from the
		most unstable mode $m=1$. For the SGWB from ultralight scalar bosons,
		since the most unstable mode already has a typically long GW emission
		timescale, the contribution of higher modes to the total
		$\Omega_\mathrm{GW}(f)$ is in general much smaller, and therefore the
		contribution from these modes was not considered
		in Refs.~\cite{brito_short,Tsukada:2019}. Ultralight vector bosons, on the
		other hand, exhibit much smaller instability timescales and therefore
		one might expect that the contribution from $m>1$ modes could be
		important.
	\begin{figure}
		\centering
		\includegraphics[width=\linewidth]{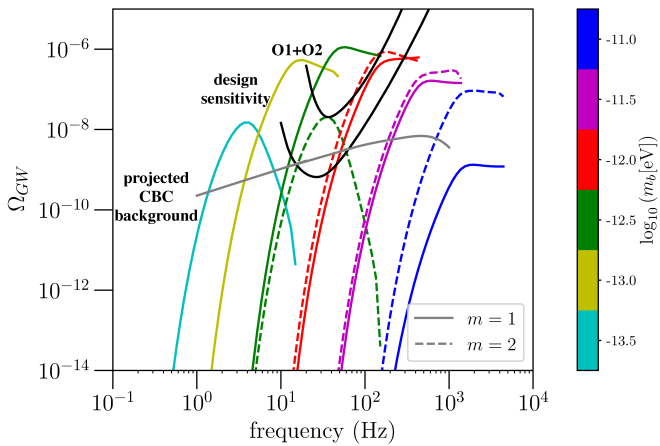}
		\caption{\label{fig:omega_gw_PIs} Contribution to the energy density
		spectra of different $m$ modes. The solid curves show the
		contribution from the $m=1$ mode for the isolated BH model, whereas
		the dashed curves represent the contribution from the $m=2$ mode. For
		the BH spin, we assume an uniform distribution $ \chi \in [0, 0.8] $.
		The black solid curves are the ($2\sigma$) power-law integrated
		sensitivity curves \cite{PI_curve}, obtained using LIGO's first (O1)
		and second (O2) observing runs \cite{LVC_O2iso}, and for Advanced LIGO
		at design sensitivity \cite{PI_design}. For comparison, we also show
		the predicted CBC background\cite{GW170817_implication} (gray solid
		curve), which is extrapolated down to \SI[]{1}{Hz} using a power-law spectrum model.}
	\end{figure}

	To study the impact of modes with $m>1$, in Fig.~\ref{fig:omega_gw_PIs} we compare the background produced by the $m=1$ (solid lines) and $m=2$ (dashed lines) modes, considering only the isolated BH channel, and assuming a uniform distribution $ \chi \in [0, 0.8] $ for the natal BH spin. For this BH population, one can see that the contribution from the $m=2$ mode is generally small for $m_b\lesssim 10^{-12}$ eV, but can be as important as, or even dominate over, the contribution from the $m=1$ mode for $m_b\gtrsim 10^{-12}$ eV.
	To understand why this happens, we note that, for a given BH mass and mode $m$, the critical BH spin below which a given mode is stable [Eq.~\eqref{eq:synchcondition}] increases with the boson mass, whereas for a fixed $M\mu$ it decreases with $m$.
	 For example, for $m_b\sim \SI{e-11}{\electronvolt}$, a majority of the BH population is only unstable against $m>1$ modes. For the few BHs that spin sufficiently fast to be unstable against the $m=1$ mode, their natal spin is close to the critical spin, and hence much less energy is extracted and emitted by the $m=1$ mode compared to the $m=2$ mode.

	 We note, however, that in the band where LIGO is most sensitive, the $m=1$ contribution is in general dominant (see Fig.~\ref{fig:omega_gw_PIs}), and the contribution from the $m>2$ modes is expected to be much smaller. Therefore we will only include the contribution of the $m=1$ and $m=2$ modes in the signal model used in Secs.~\ref{sec:5} and \ref{sec:6}, where we show the results of injection studies and a search of this signal in Advanced LIGO's data.

\section{\label{sec:4}Search Method}
	In this section, we review the conventional search method and Bayesian statistics framework we use in order to either claim a detection, or to place constraints on the SGWB model we presented above, when searching for such a background in LIGO data.

	\subsection{Definitions}
	For a single baseline with a pair of detectors, the SGWB is analyzed using the cross-correlation between two output streams. Although the formalism can be extended to handle a larger network of detectors \cite{allen_joe,joe_neil}, we will consider this simpler case as we only analyze the data from the two LIGO detectors. Following the notation in Ref.~\cite{tom_nonGR_method}, we define a cross-correlation estimator that is optimal for a Gaussian background as \cite{sgwb_hanford, joe_neil}
	\begin{align}
		\label{eq:estimator_def}
		\hat{C}(f) \equiv \frac{f^3}{T}\frac{20\pi^2}{3H_0^2}\tilde{s}_1^*(f)\tilde{s}_2(f) \ .
	\end{align}
	Here, $ \tilde{s}_i(f) $ is the Fourier transform of the time series output of the $ i $-th detector, $ T $ is the total observation time, and $H_0$ is the Hubble paramater. This is normalized such that
	\begin{align}
		\label{eq:estimator_mean}
		\left\langle \hat{C}(f)\right\rangle =\gamma(f) \Omega_\mathrm{GW}(f),
	\end{align}
	where $ \gamma(f) $ is the overlap reduction function \cite{ORF}.
	In the low signal-to-noise ratio limit, the variance of the cross-correlation estimator $\hat{C}(f)$ is approximately given by
	\begin{align}
		\label{eq:sigma_method}
		\sigma^2(f)\approx\frac{1}{2T\Delta f}\left(\frac{10\pi^2f^3}{3H_0^2}\right)^2P_1(f)P_2(f),
	\end{align}
	where $ \Delta f $ is the frequency resolution and $ P_i(f) $ is the power spectral density (PSD) of the $ i $-th detector.

	\subsection{\label{sec:bayes}Bayesian inference}
	Following the method in Ref.~\cite{PE_stochastic}, we discuss a Bayesian formalism for our detection statistics, parameter estimation, and model selection. For our analysis, Bayes' theorem states that, using the estimator $\hat{C}(f)$,
	\begin{align}
		\label{eq:bayes_post}
		p(\boldsymbol{\theta}_\mathcal{A}|\{\hat{C}\}, \mathcal{A})=\frac{L(\{\hat{C}\}|\boldsymbol{\theta}_\mathcal{A}, \mathcal{A})\pi(\boldsymbol{\theta}_\mathcal{A}|\mathcal{A})}{Z(\{\hat{C}\}|\mathcal{A})},
	\end{align}
	where $p(\boldsymbol{\theta}_\mathcal{A}|\{\hat{C}\}, \mathcal{A})$ is the posterior probability on the multi-dimensional space of parameters $ \boldsymbol{\theta}_\mathcal{A} $ that describe the SGWB model in the signal hypothesis $ \mathcal{A} $, $L(\{\hat{C}\}|\boldsymbol{\theta}_\mathcal{A}, \mathcal{A})$ is the likelihood,  $\pi(\boldsymbol{\theta}_\mathcal{A}|\mathcal{A})$ is the prior probability of the parameters $\boldsymbol{\theta}_\mathcal{A}$, and $Z(\{\hat{C}\}|\mathcal{A})$ is the evidence. We use a nested sampling package \texttt{PyMultiNest} \cite{pymultinest} to evaluate the likelihood. \texttt{PyMultiNest} is a python interface to the nested sampling package \texttt{MultiNest} \cite{multinest1, multinest2, multinest3}, which produces a set of samples drawn from an estimated posterior.

	Let $ \{\hat{C}\} $ be the cross-correlation estimator obtained from the data within an analyzed frequency band. For a given $ \{\hat{C}\} $, we define a Gaussian likelihood for every frequency bin, and hence a joint likelihood given by the product of each likelihood, such that
	\begin{widetext}
		\begin{align}
			 \ln \left[ L(\{\hat{C}\}|\boldsymbol{\theta}_\mathcal{A}, \mathcal{A})\right]&=\sum_{f} \ln \left[L(\hat{C}(f)|\boldsymbol{\theta}_\mathcal{A}, \mathcal{A})\right]\nonumber\\
			\label{eq:L_f}
			&=\sum_f \left\{-\frac{\left[\hat{C}(f)-\gamma(f)\Omega_\mathcal{A}(f;\boldsymbol{\theta}_\mathcal{A})\right]^2}{2\sigma^2(f)}-\frac{1}{2}\ln\left(2\pi\sigma^2(f)\right)\right\}.
		\end{align}
	\end{widetext}
	Here, $ \Omega_\mathcal{A}(f;\boldsymbol{\theta}_\mathcal{A}) $ is a model energy-density spectrum for a given set of parameters $ \boldsymbol{\theta}_\mathcal{A} $.

        For the priors, we set a log-uniform prior\footnote{At this point we
        choose to adopt mostly uninformative priors for
        $\boldsymbol{\theta}_\mathcal{A}$, since this is the most
        conservative choice. However, in principle, we could also choose a
        prior that encodes prior knowledge on the boson mass obtained from
        independent experiments, such as constraints from BH spin
        measurements in X-ray
        binaries~\cite{sr_vector_4,Cardoso:2018tly,superradiance,Stott:2020gjj}.
        This will be revisited in future work.} on the vector mass $m_b$, and
        a linearly uniform prior on the BH spin upper/lower limits
        $\chi_{\mathrm{ul}/\mathrm{ll}}$. Therefore, following Bayes' theorem
        [Eq.~\eqref{eq:bayes_post}], the posterior probability is inversely
        proportional to the boson mass:
	\begin{align}
		p(\boldsymbol{\theta}_\mathcal{A}|\{\hat{C}\}, \mathcal{A}) \propto \frac{1}{m_b} L(\{\hat{C}\}|\boldsymbol{\theta}_\mathcal{A}, \mathcal{A}).
	\end{align}

	We will also be interested in performing model selection between
	different signal models. The Bayesian evidence for a given hypothesis
	quantifies how well the model fits the obtained data and is defined as
	\begin{align}
		Z(\{\hat{C}\}|\mathcal{A})=\int L(\{\hat{C}\}|\boldsymbol{\theta}_\mathcal{A}, \mathcal{A})\pi(\boldsymbol{\theta}_\mathcal{A}|\mathcal{A})\mathrm{d}^D\boldsymbol{\theta}_\mathcal{A} \ .
	\end{align}
	This expression can also be interpreted as the fully-marginalized likelihood over the entire parameter space. In the case where no signal is present (the null hypothesis), the evidence is obtained by fixing $\Omega_{\mathcal{A}}(f;\theta_\mathcal{A})$ to zero in the likelihood [Eq.~\eqref{eq:L_f}].
	To assess which hypothesis, $ \mathcal{A}$ or $\mathcal{B}$, better describes the observed data, we can  compute the odds ratio $\mathcal{O}^{\mathcal{A}}_{\mathcal{B}}$ defined as
	\begin{align}
		\mathcal{O}_\mathcal{B}^\mathcal{A}\equiv\frac{p(\mathcal{A}|\{\hat{C}\})}{p(\mathcal{B}|\{\hat{C}\})}=\frac{Z(\{\hat{C}\}|\mathcal{A})}{Z(\{\hat{C}\}|\mathcal{B})}\frac{\pi(\mathcal{A})}{\pi(\mathcal{B})} \ ,
		\label{eq:ms_BF}
	\end{align}
	where $ Z(\{\hat{C}\}|\mathcal{A})$ and $Z(\{\hat{C}\}|\mathcal{B}) $ are
	the evidences for the hypotheses $\mathcal{A}$ and $\mathcal{B}$
	respectively, whereas $\pi(\mathcal{A})$ and $\pi(\mathcal{B})$ are the
	prior probability of the respective hypothesis. Hereafter, we will set
	the \textit{a priori} probability ratio for the two models,
	$\pi(\mathcal{A})/\pi(\mathcal{B})$, to unity. Therefore, for our case,
	the odds ratio will be effectively equivalent to the Bayes factor
	(defined as the ratio between the evidences). In what follows, we will
	evaluate the statistical significance for a given hypothesis in terms of
	the Bayes factor and follow the convention that a natural logarithmic
	Bayes factor $\approx 8$ indicates that one model is favored over the
	other with great confidence \cite{Jeffreys}.

\section{\label{sec:5}Results}

	We are now in a position to study the sensitivity of Advanced LIGO to
	the SGWB we described in Sec.~\ref{sec:3}, using the tools introduced
	in the previous section. We first study the range of vector boson
	masses that LIGO will be able to probe at its design sensitivity by
	performing injections of the SGWB model into synthesized data,
	and then explore if one could successfully discriminate between the
	SGWB due to CBCs, and the one due to the superradiant instability.

	\subsection{\label{sec:sw} Range of sensitivity for vector boson masses}

	The results shown in Fig.~\ref{fig:omega_gw_PIs} suggest that, for vector bosons with masses roughly in the range $m_b\sim [10^{-13}, 10^{-12}]$ eV, the SGWB could be detected by LIGO at design sensitivity, and that even with the sensitivity of LIGO's first and second observing runs, one could already probe bosons with masses $m_b\sim 10^{-12.5}$ eV.
	To study this in more detail, and assess the vector mass range that we
	can probe through this method, we inject our SGWB model into synthesized
	data with different values of the model parameters, and then
	infer the detectability of the signal by computing Bayes factors between
	the signal and noise hypotheses.

	We follow the injection scheme described in Ref.~\cite{Tsukada:2019}, where the simulated cross-correlation spectrum is defined as
	\begin{align}
		\label{eq:C_inj}
		\hat{C}_\mathrm{sim}(f)\equiv\gamma_{\mathrm{HL}}(f)\Omega_{\mathrm{inj}}(f;\boldsymbol{\theta}_\mathrm{inj})+\sigma(f)\hat{n} \ .
	\end{align}
	Here, $ \gamma_{\mathrm{HL}}(f)\ $ is the overlap reduction function for
	the LIGO baseline \cite{ORF}, $
	\Omega_\mathrm{inj}(f;\boldsymbol{\theta}_\mathrm{inj}) $ is the injected
	background for the given model parameters $
	\boldsymbol{\theta}_\mathrm{inj} $, and $ \hat{n} $ is a random variable
	drawn from a Gaussian distribution with zero mean and unit variance. To
	synthesize data, we set $\Delta f$ as \SI[]{0.25}{Hz} for the injection studies in this and next subsections, and specify $\sigma(f)$ using Eq.~\eqref{eq:sigma_method}
	for a given PSD and observation time. We assume 3 years of observation
	with the projected design sensitivity PSD for the LIGO
	detectors~\cite{PI_design} (see Fig.~\ref{fig:omega_gw_PIs}). From this
	simulated cross-correlation spectrum, our pipeline computes the
	likelihood function and the posterior distribution of the parameters $
	\boldsymbol{\theta}_\mathrm{inj} $, as well as the Bayesian evidence
	$Z(\{C_\mathrm{inj}\})$.
	\begin{figure}[h]
		\centering
		\includegraphics[width=\linewidth]{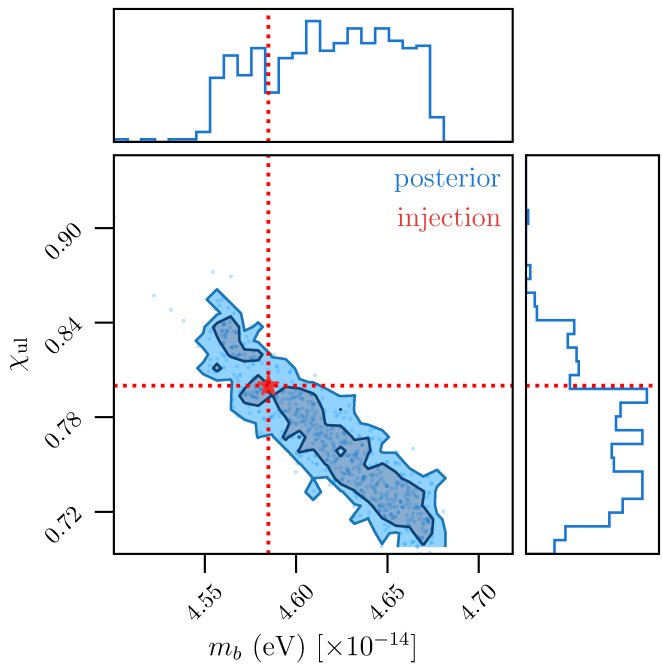}
		\caption{The posterior samples recovered for one of the injections
		$\Omega_\mathrm{inj}(f; \boldsymbol{\theta})$ where the injected
		parameters $\boldsymbol{\theta}_\mathrm{inj}$ are $(m_b,
		\chi_\mathrm{ul})=(\SI{4.58e-14}{eV}, 0.8)$, and $\chi_\mathrm{ll}=0$
		is kept fixed in the parameter recovery. The star marker indicates
		the true values for the injected parameters. The contours represent
		the 2 and 3--$\sigma$ credible regions.}
		\label{fig:posterior_inj}
	\end{figure}

	In Fig.~\ref{fig:posterior_inj}, we show an example of the parameter
	estimation results for one of the injections, where the injected
	parameters $\boldsymbol{\theta}_\mathrm{inj}$ are $m_b =
	\SI{4.58e-14}{eV}$ and $(\chi_\mathrm{ul}, \chi_\mathrm{ll})=(0.8, 0)$.
	In the parameter estimation recovery, we adopt the more conservative
	parametrization for $p(\chi)$ where $\chi_\mathrm{ll}=0$ is kept fixed
	while $\chi_\mathrm{ul}$ is allowed to vary. In order to obtain evidence
	for the noise hypothesis, we also perform the same analysis for the
	identical noise realization without the injection. The star marker
	represents the true parameters of the injection, which lie within the 2
	and 3--$\sigma$ contours. For this recovered injection, we estimate the
	signal-to-noise ratio to be 20.2, and find that the log Bayes factor of
	the signal versus noise hypotheses is approximately 500, showing that the
	signal is detected with great confidence.

\begin{figure}[h]
		\centering
		\includegraphics[width=\linewidth]{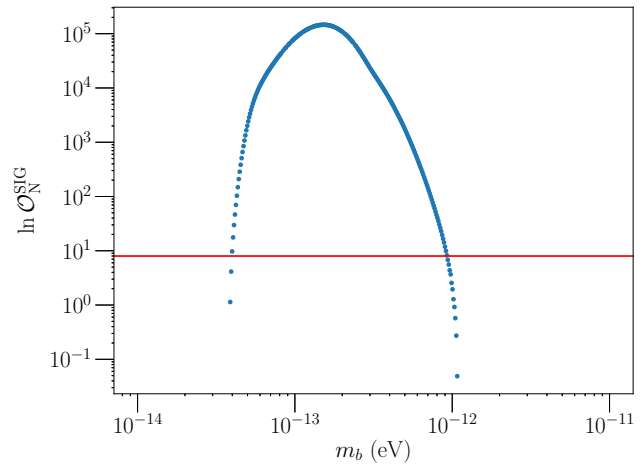}
		\caption{The Bayes factor of the recovered signal as a function of injected $m_b[\mathrm{eV}]$, fixing $\chi_\mathrm{ll}=0$ and $\chi_\mathrm{ul}=0.8$ in the injected signals. The red horizontal line shows our detection criterion $\ln \mathcal{O}^\mathrm{SIG}_\mathrm{N}=8$. Injections that have a Bayes factor above the red line are confidently detected.}
		\label{fig:sense_window}
	\end{figure}

	We repeat this injection recovery varying the vector boson mass of each
	injection, but using the same BH spin distribution for all injections,
	namely a uniform distribution with $\chi_\mathrm{ll}=0$ and
	$\chi_\mathrm{ul}=0.8$ [see Eq.~\eqref{eq:spin_dist}]. Our results are
	summarized in Fig.~\ref{fig:sense_window}, where we show the log Bayes
	factor as a function of the vector boson mass for this set of injections.
	These results indicate that, given the detection criteria of $\ln
	\mathcal{O}^\mathrm{SIG}_\mathrm{N}=8$, we could detect vector
	bosons with masses in the range
	\SIrange[]{5e-14}{e-12}{\electronvolt}. We note that this detectable
	mass range is wider than that for ultralight scalar bosons due
	to the enhanced energy spectrum for masses below $m_b\lesssim
	\SI[]{e-13}{\electronvolt}$ (cf. Fig. 5 in Ref.~\cite{Tsukada:2019}).

	\subsection{\label{sec:ms} Distinguishing the background from vector clouds and CBCs}
	In the previous subsection, we neglected the fact that besides the SGWB
	signal due to the formation of vector clouds, CBCs also produce a
	background that is expected to be detectable by Advanced LIGO at design
	sensitivity, and so in reality we should consider \textit{both} types of SGWBs in our simulation.
	Fig.~\ref{fig:omega_gw_PIs} illustrates that the SGWB signal from the
	vector clouds can dominate over the projected CBC background
	for some choices of the vector mass and BH spin distribution, and
	therefore a natural question to ask is whether we can distinguish between
	these two signal models based on the Bayesian framework of
	Sec.~\ref{sec:ms}. A similar study was done in Ref.~\cite{Tsukada:2019}
	for the case of scalar boson clouds, and here we repeat this study for
	the vector boson model.

	We consider an energy density spectrum that consists of the contributions
	from both the vector cloud and the projected CBC background, which
	reads
	\begin{align}
		\label{eq:ms_sgwb_model}
		\Omega_\mathrm{inj}(f; \boldsymbol{\theta}) = \Omega_\mathrm{inj}^\mathrm{VC}(f; \boldsymbol{\theta}) + \Omega_\mathrm{inj}^\mathrm{CBC}(f).
	\end{align}
	$\Omega_\mathrm{inj}^\mathrm{VC}(f; \boldsymbol{\theta})$ is the
	background due to superradiant instabilities under a $\chi_\mathrm{ul}$ parametrization of
	the natal spin distribution for isolated BHs [Eq.~\eqref{eq:SI_sgwb}], and
	$\Omega_\mathrm{inj}^\mathrm{CBC}(f) $ is the fixed CBC background
	approximated as a power-law spectrum,
	\begin{align}
		\label{eq:omega_CBC}
		\Omega_\mathrm{GW}^\mathrm{CBC}(f) = 1.8\times10^{-9}\left(\frac{f}{25\mathrm{Hz}}\right)^{2/3},
	\end{align}
	as inferred from \cite{GW170817_implication}.

	Due to the computational expense, for the injection recovery of the vector cloud
	background, we only consider the isolated BH channel,
	[Eq.~\eqref{eq:iso_omega_gw}]. Since the merger remnant channel is
	subdominant for the boson masses of interest (see
	Fig.~\ref{fig:omega_gw_channels}), and slightly below the projected CBC
	background, we do not expect the inclusion of this additional channel to
	change the results significantly. Like we did in the previous subsection,
	we adopt the $\chi_\mathrm{ul}$ parametrization for $p(\chi)$, i.e. we
	use a parametrization where $\chi_\mathrm{ll}=0$ is kept fixed while
	$\chi_\mathrm{ul}$ is allowed to vary. The injected CBC background is
	recovered with the following parametrization\footnote{We refer to
	Eq.~\eqref{eq:CBC_rec} as a CBC model, even though this could be
	generally called a ``power-law spectrum model.'' It has been shown that the systematic error potentially caused by this bias is below the statistical error and hence would not affect the detectability of the background \cite{Callister:2016aa, Saffer:2020aa}.}
	\begin{align}
		\Omega_{\mathrm{rec}}^\mathrm{CBC}(f;\Omega_0, \alpha)\equiv \Omega_0\left(\frac{f}{25\mathrm{Hz}}\right) ^\alpha \ .
		\label{eq:CBC_rec}
	\end{align}

\begin{figure}[h]
		\centering
		\includegraphics[width=\linewidth]{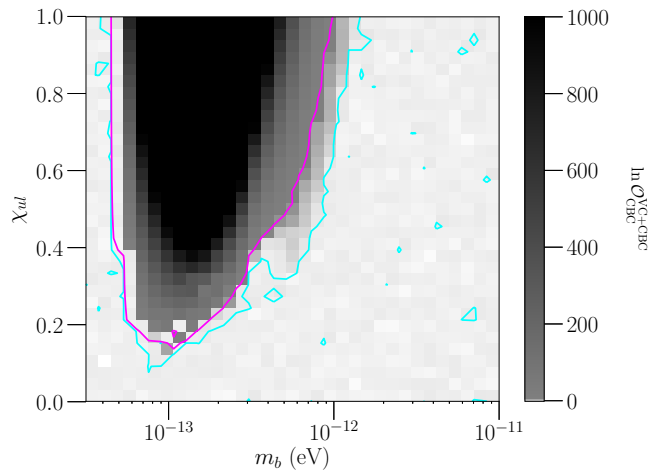}
		\caption{Gray scale map of a log Bayes factor between CBC-only and
		the joint VC$+$CBC models with two contours of $ \ln
		\mathcal{O}^\mathrm{VC+CBC}_\mathrm{CBC}=8$ (magenta) and $ 0 $
		(cyan). The parameter space inside the magenta contour indicates the
		region where one can confidently discern the vector cloud background
		from the projected CBC background.}
		\label{fig:ms_BF}
	\end{figure}

	To see whether we can detect the vector cloud background in the presence
	of the CBC background, we will compute a log Bayes factor between two
	hypotheses: a CBC-only hypothesis, corresponding to the hypothesis that
	only the CBC background is present in the data, and a joint vector cloud
	and CBC hypothesis (VC+CBC), corresponding to the hypothesis that both
	the CBC and vector cloud backgrounds are present in the data. The
	parameters considered when evaluating the evidence for each background
	model are listed in Table~\ref{tab:parameters}. We repeat this
	computation for several injections, varying the parameters $ (m_b,
	\chi_\mathrm{ul}) $ until we explore a grid over the entire prior space.
	The results of this study are shown in Fig.~\ref{fig:ms_BF}, where we
	plot a gray-scale map of the log Bayes factors, highlighting the contours
	where $ \ln \mathcal{O}^\mathrm{VC+CBC}_\mathrm{CBC}=8$ (magenta contour)
	and $\ln \mathcal{O}^\mathrm{VC+CBC}_\mathrm{CBC}=0 $ (cyan contour). The
	parameter space inside the magenta contour indicates the region where one
	can confidently discern the vector cloud background from the projected
	CBC background. As expected, for large $\chi_\mathrm{ul}$ the range of
	boson masses for which one could claim a confident detection agrees with
	the one obtained in the previous subsection.
	\begin{table}[h]
		\begin{ruledtabular}
			\begin{tabular}{c|cc}
			Models & CBC-only & VC$+$CBC \\ 
			\hline
			Parameters & $\Omega_0, \alpha$ & $m_b, \chi_\mathrm{ul}, \Omega_0, \alpha$
			\end{tabular}
			\caption{\label{tab:parameters}Parameters in each recovered background model.}
		\end{ruledtabular}
	\end{table}

\section{\label{sec:O1O2_zerolag} Constraints on ultralight vector bosons using LIGO data}
	
	Using the Bayesian framework presented in Sec.~\ref{sec:4}, we now
	conduct a search for the vector cloud background
	[Eq.~\eqref{eq:omega_gw}] using the cross-correlation spectra obtained
	from the first (O1) and second (O2) observing runs of Advanced LIGO
	\cite{LVC_O2iso}. (The data products used in this paper are publicly
	available at Ref.~\cite{O1O2_data}.) The analysis is conducted following
	the prescription presented in the injection studies done in
	Secs.~\ref{sec:sw} and \ref{sec:ms}, except for the likelihood
	evaluation. Since in this case the independent cross-correlation spectrum
	is obtained from each observing run, the full likelihood expression takes
	the form
	\begin{align}
		L(\hat{C}_\mathrm{O1}, \hat{C}_\mathrm{O2} | \boldsymbol{\theta}, \mathcal{H}_\mathrm{VC}) = L(\hat{C}_\mathrm{O1} | \boldsymbol{\theta}, \mathcal{H}_\mathrm{VC})L(\hat{C}_\mathrm{O2} | \boldsymbol{\theta}, \mathcal{H}_\mathrm{VC})\,,
	\end{align}
	where each likelihood in the right hand side follows the definition of
	Eq.~\eqref{eq:L_f}. These cross-correlation spectra are provided over the frequency range from \SIrange[]{20}{700}{Hz} with the
	frequency resolution of $\Delta f=\SI[]{1/32}{Hz}$.

\begin{figure}
		\centering
		\includegraphics[width=\linewidth]{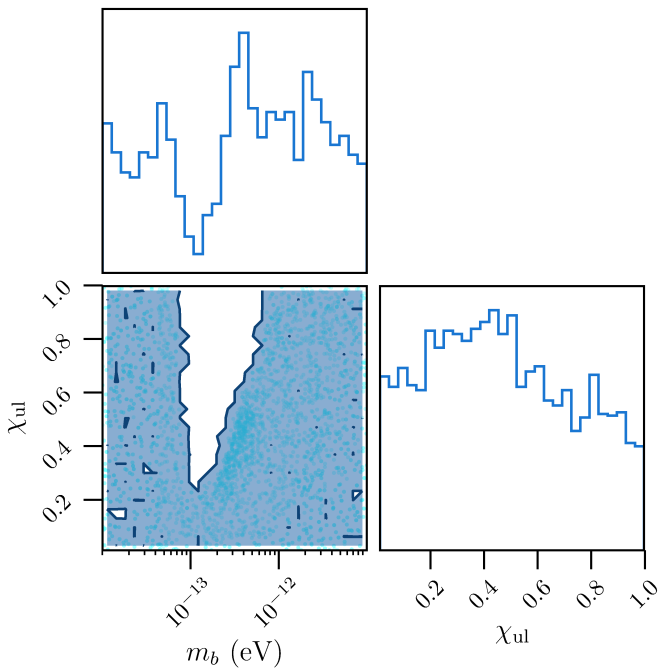}
		\caption{\label{fig:O1O2_post_ul} Posterior results obtained with the data from the first and second observing runs of Advanced LIGO, recovered with the $ \chi_\mathrm{ul}$ parameterization. The contour on the two-dimensional posterior represents the 95\% confidence level.}
	\end{figure}

\begin{figure}
		\centering
		\includegraphics[width=\linewidth]{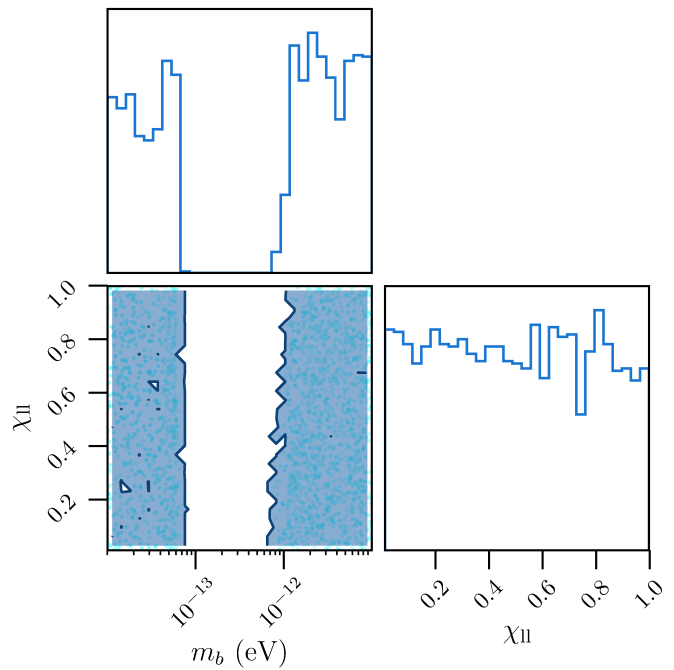}
		\caption{\label{fig:O1O2_post_ll} Posterior results, analogous to Fig.~\ref{fig:O1O2_post_ul}, for the $\chi_\mathrm{ll}$ spin parameterization.}
	\end{figure}

	We do not find statistically significant evidence for a vector cloud
	background. Therefore, we place constraints on the two dimensional space
	($m_b,\chi_{\mathrm{ul},\mathrm{ll}}$) using the estimated posterior
	probability distribution. Figures~\ref{fig:O1O2_post_ul} and
	\ref{fig:O1O2_post_ll} show the posteriors under the $\chi_\mathrm{ul}$
	and $\chi_\mathrm{ll}$ parametrizations, respectively [see
	Eq.~\eqref{eq:spin_dist} and corresponding discussion]. In particular,
	the results shown in Fig.~\ref{fig:O1O2_post_ul} indicate that, when
	using the $\chi_\mathrm{ul}$ parametrization, the data disfavors boson
	masses close to $m_b\approx\SI{e-13}{\electronvolt}$ and relatively high
	$\chi_\mathrm{ul}\gtrsim 0.2$. We note, however, that the marginalized 1D
	posterior for $m_b$ does not indicate a strong constraint at the 95\%
	confidence level. On the other hand, when we fix $\chi_\mathrm{ul}=1$ and
	allow the lower bound $\chi_\mathrm{ll}$ to vary, then
	Fig.~\ref{fig:O1O2_post_ll} suggests that the mass range
	\SIrange{0.8e-13}{6.5e-13}{\electronvolt} is excluded (95\% credible
	interval) regardless of the spin's lower bound $\chi_\mathrm{ll}$, as can
	be seen in the 1D marginalized posterior of $m_b$.
	
	 In summary, these results suggest that minimally coupled vector bosons
	 with masses around $m_b\approx\SI{e-13}{\electronvolt}$ are
	 highly disfavored, unless most stellar-mass BHs are born with a small
	 spin.

\section{\label{sec:6}Conclusion}
   	In this paper, we computed and studied in detail the SGWB produced by
   	the superposition of GW signals from extragalactic BH-ultralight
   	vector cloud systems formed through the superradiant instability.
   	Using a Bayesian framework, we performed the first search for such
   	signal in LIGO data. This extends previous
   	works~\cite{brito_short,Tsukada:2019} where a similar background for
   	scalar bosons was studied. We also improved on those works by adding
   	the contribution of the second most unstable mode, $m=2$ (in addition
   	to the most unstable one, $m=1$) to the SGWB model, which was not
   	considered in Refs.~\cite{brito_short,Tsukada:2019}. In particular, we
   	found that the contribution from the $m=2$ mode can be as important
   	as, or even dominate over, the $m=1$ mode for $m_b \geq
   	\SI{e-12}{\electronvolt}$, and therefore affect the detectability and
   	potential constraints on the vector boson mass.

	To estimate the potential detectable window, we performed injection tests
	on simulated Advanced LIGO data at design sensitivity, and found that
	Advanced LIGO is especially sensitive to the background emitted by
	minimally coupled vector bosons with masses in the range $\sim [5\times
	10^{-14},10^{-12}]$ eV (see Fig.~\ref{fig:sense_window}). This detectable
	mass range is broader than that for ultralight scalar bosons (cf. Fig 5
	in Ref.~\cite{Tsukada:2019}), especially at small masses, due to the
	considerably larger GW power for vector bosons. We also studied the
	capability to claim a detection for this background model in the presence
	of a (fiducial) CBC background model. Our results suggest that we can
	distinguish between both models in a large part of the parameter space
	(see Fig.~\ref{fig:ms_BF}).
	
	Additionally, in the future we may be able to place constraints using
	only BBH merger remnants, which are much less sensitive to unknown BH
	population statistics. We performed similar simulations using the LIGO's
	design sensitivity with and without an injection to search for the BBH
	remnant component alone. With the injection of the SGWB for
	$m_b=\SI[]{e-13}{eV}$, we recovered it with a Bayes factor of 10.4 and
	consistent vector mass estimation. On the other hand, without the
	injection, we ruled out $m_b\sim \SI[]{e-13}{eV}$ at the 95\% confidence
	level. Although we leave the model selection test between this and the
	CBC background as future work, these results indicate that we will
	potentially be able to make a detection, or place more robust constraints
	on the vector mass, by probing the BBH remnant signal model.

	Lastly, we presented results obtained by analyzing data from Advanced
	LIGO's first and second observing runs. We did not find any signal
	consistent with our vector cloud model, independent of the
	parametrizations employed for the BH spin distribution. For the more
	optimistic parametrization (the $\chi_\mathrm{ul}$ parametrization), we
	rule out (minimally coupled) ultralight vector bosons in the mass range
	\SIrange{0.8e-13}{6.5e-13}{\electronvolt} with 95\% percent credibility.
	A less optimistic parametrization $\chi_\mathrm{ll}$, which allows for
	the possibility that all isolated BHs have negligible spins, does not
	give strong constraints. However, boson masses around $\sim 10^{-13}$ eV
	are highly disfavored by our results, unless all isolated BHs form with
	spins $\lesssim 0.2$. We note that these constraints depend on our
	specific choice for the BH spin distribution, as well as the
	astrophysical models we adopted in this analysis.

	Aside from these constraints, the observation of stellar-mass BHs in
	X-ray binaries spinning above the superradiant threshold already
	disfavors the existence of ultralight vector fields in the range $\sim
	[10^{-14}, 10^{-11}]\SI[]{}{\electronvolt}$
	\cite{sr_vector_4,Cardoso:2018tly,superradiance,Stott:2020gjj}, which
	overlaps with the range of masses we are able to probe with the SGWB.
	However, we should note that such constraints should be interpreted with
	caution, since BH spin measurements from X-ray binaries are often
	susceptible to large systematic uncertainties (see e.g.
	Ref.~\cite{Krawczynski:2018fnw}). GW searches should therefore be seen as
	complementary to those measurements. Given sufficiently robust estimates
	of all the relevant uncertainties, it could be interesting to include GW
	searches and BH spin measurements in the same Bayesian framework, which
	would in principle lead to stronger constraints.

 	We only considered a minimally coupled ultralight vector field,
 	neglecting possible (non-gravitational) couplings with other particles,
 	as well as any non-trivial self-interactions beyond the mass term. Our
 	results apply to any massive vector field as long as any additional
 	interactions are negligible compared to the gravitational interaction
 	between the BH and vector field. Sufficiently large non-gravitational
 	interactions could change the picture, in particular by affecting the
 	evolution of the superradiant instability. For example, for ultralight
 	scalar fields it has been shown that
 	self-interactions~\cite{Arvanitaki_precision,Yoshino:2012kn,Yoshino:2015nsa}
 	and couplings to
 	photons~\cite{Rosa:2017ury,Boskovic:2018lkj,Ikeda:2018nhb} can quench
 	the superradiant instability, and effectively increase the timescale
 	needed to extract a substantial amount of energy and angular momentum
 	from the BH~\cite{Fukuda:2019ewf,Mathur:2020aqv}. The effect of such
 	interactions has been less studied for massive vector fields---with the
 	exception of some consideration of the case where the vector boson mass
 	arises through a Higgs mechanism~\cite{sr_vector_4,Fukuda:2019ewf}---but
 	one might expect that similar results also apply for this case. It would
 	be important to study this question in more detail in the future.
 	Finally, we should note that if ultralight dark matter photons couple
 	directly to ordinary matter, they could also produce another kind of
 	observable signal in GW interferometers by inducing displacements on the
 	LIGO
 	mirrors~\cite{Pierce:2018xmy,Guo:2019ker,Miller:2020vsl,Michimura:2020vxn}.
 	Since our results mainly apply to ultralight bosons for which
 	non-gravitational interactions are negligible, our constraints
 	complement those searches.

\begin{acknowledgments}
	L.T is also grateful for computational resources provided by the LIGO lab
	and supported by National Science Foundation Grants PHY-0757058 and
	PHY-0823459. L.T. is supported by JSPS KAKENHI Grant Number JP18J21709.
	N.S. thanks Masha Baryakhtar for useful discussions during the completion
	of this work. N.S. and W.E. acknowledge support from an NSERC Discovery
	grant. Research at Perimeter Institute is supported in part by the
	Government of Canada through the Department of Innovation, Science and
	Economic Development Canada and by the Province of Ontario through the
	Ministry of Colleges and Universities. R.B. acknowledges financial
	support from the European Union's Horizon 2020 research and innovation
	programme under the Marie Sk\l odowska-Curie grant agreement No. 792862, 
	from the European Union's H2020 ERC, Starting Grant agreement no.~DarkGRA--757480, 
	and from the MIUR PRIN and FARE programmes (GW-NEXT, CUP:~B84I20000100001).
	This research was supported by the Amaldi Research Center funded by the
	MIUR program ``Dipartimento di Eccellenza'' (CUP:~B81I18001170001).
\end{acknowledgments}

\bibliography{bibliography}

\end{document}